\documentstyle[12pt,twoside]{article}
\input psfig.sty
\def\double12 {\smallskipamount=6pt plus2pt minus2pt
                  \medskipamount=12pt plus4pt minus4pt
                  \bigskipamount=24pt plus8pt minus8pt
                  \normalbaselineskip=24pt plus0pt minus0pt
                  \normallineskip=2pt
                  \normallineskiplimit=0pt
                  \jot=6pt
                  {\def\smallskip {\vskip\smallskipamount}}
                  {\def\medskip   {\vskip\medskipamount}}
                  {\def\bigskip   {\vskip\bigskipamount}}
                  {\setbox\strutbox=\hbox{\vrule
                    height17.0pt depth7.0pt width 0pt}}
                  \parskip 0pt
                  \normalbaselines}

\def\half12 {\smallskipamount=6pt plus2pt minus2pt
                  \medskipamount=12pt plus4pt minus4pt
                  \bigskipamount=24pt plus8pt minus8pt
                  \normalbaselineskip=16pt plus0pt minus0pt
                  \normallineskip=2pt
                  \normallineskiplimit=0pt
                  \jot=6pt
                  {\def\smallskip {\vskip\smallskipamount}}
                  {\def\medskip   {\vskip\medskipamount}}
                  {\def\bigskip   {\vskip\bigskipamount}}
                  {\setbox\strutbox=\hbox{\vrule
                    height17.0pt depth7.0pt width 0pt}}
                  \parskip 0pt
                  \normalbaselines}

\def\single12 {\smallskipamount=3pt plus2pt minus2pt
                  \medskipamount=6pt plus4pt minus4pt
                  \bigskipamount=12pt plus8pt minus8pt
                  \normalbaselineskip=12pt plus0pt minus0pt
                  \normallineskip=1pt
                  \normallineskiplimit=0pt
                  \jot=3pt
                  {\def\smallskip {\vskip\smallskipamount}}
                  {\def\medskip   {\vskip\medskipamount}}
                  {\def\bigskip   {\vskip\bigskipamount}}
                  {\setbox\strutbox=\hbox{\vrule
                    height8.5pt depth3.5pt width 0pt}}
                  \parskip 0pt
                  \normalbaselines}

\def\refitem{\par\noindent\hangindent 20pt}

\def\wisk#1{\ifmmode{#1}\else{$#1$}\fi}

\def\deg    {\wisk{^\circ}}
\def\ddeg   {\wisk{{\rlap.}^\circ}}

\begin{document}
\pagestyle{plain}
\half12

\large
\begin{center}
Spatial Correlation Between H$\alpha$ Emission and Infrared Cirrus
\end{center}

\medskip
\normalsize
\half12
\noindent
\begin{center}
A.~Kogut\footnotemark[1]$^{,2}$
\end{center}
\footnotetext[1]{
~Hughes STX Corporation, Laboratory for Astronomy and Solar Physics, 
Code 685, NASA/GSFC, Greenbelt MD 20771. \newline
\indent~$^2$ E-mail: kogut@stars.gsfc.nasa.gov. \newline
}

\medskip
\normalsize
\half12
\begin{center}
Accepted for publication by
{\it The Astronomical Journal} \\
June 17, 1997\\
\end{center}


\medskip
\begin{center}
\large
ABSTRACT
\end{center}

\normalsize
\noindent
Cross-correlation of the DIRBE 100 $\mu$m survey
with previously published H$\alpha$ maps 
tests correlations between far-infrared dust 
and the warm ionized interstellar medium
in different regions of the sky. 
A $10\deg \times 12\deg$ patch at Galactic latitude $b = -21\deg$
shows a correlation slope
$a_0 = 0.85 \pm 0.44$ Rayleighs MJy$^{-1}$ sr
significant at 97\% confidence.
A set of H$\alpha$ images over the north celestial polar cap
yields a weaker correlation slope
$a_0 = 0.34 \pm 0.33$ Rayleighs MJy$^{-1}$ sr.
Combined with observations from microwave anisotropy experiments,
the data show roughly similar correlations
on angular scales 0\ddeg7 to 90\deg.
Microwave experiments may observe more emission
per unit dust emission
than are traced by the same structures observed in H$\alpha$.

\clearpage
\section{Introduction}
Observations of anisotropy in the cosmic microwave background
are complicated by the presence of foreground Galactic emission
along all lines of sight.
At high latitudes ($|b| > 10\deg$),
diffuse Galactic emission is dominated by 
optically thin synchrotron, dust, and free-free emission.
In principle, these components may be distinguished by their different
spatial morphology and frequency dependence.
In practice, there is no emission component for which
both the frequency dependence and spatial distribution are well determined.
Synchrotron radiation dominates radio-frequency surveys,
but the spectral index 
steepens with frequency and
has poorly-determined spatial variation
(Bennett et al.\ 1992,
Banday \& Wolfendale 1991).
Dust emission dominates far-infrared surveys,
but its spectral behavior at longer wavelengths
depends on the shape, composition, and size distribution
of the dust grains, which are poorly known
(D\'{e}sert, Boulanger, \& Puget 1990).
Free-free emission from electron-ion interactions 
has well-determined spectral behavior
but lacks an obvious template map:
free-free emission never dominates the high-latitude radio sky,
while other tracers of the warm ionized interstellar medium (WIM)
such as H$\alpha$ emission, pulsar dispersion measure, 
or N II emission
are either incomplete, undersampled, or noise-dominated
(Bennett et al.\ 1992,
Reynolds 1992,
Bennett et al.\ 1994).

The problem is particularly acute for free-free emission:
lacking an accurate template for the spatial distribution,
estimation of free-free emission requires 
pixel-by-pixel frequency decomposition
which significantly reduces the signal to noise ratio 
of the desired cosmological signal.
Recently, 
Kogut et al.\ 1996a
proposed using infrared emission from diffuse cirrus
as a tracer of the diffuse ionized gas responsible for free-free emission.
Cross-correlation of the COBE 
Differential Microwave Radiometer (DMR)
maps at 31.5, 53, and 90 GHz 
with the Diffuse Infrared Background Experiment (DIRBE)
far-infrared maps at 100, 140, and 240 $\mu$m
shows statistically significant emission in each microwave map
whose spatial distribution 
on angular scales above 7\deg
~is traced by the far-infrared dust emission
(Kogut et al.\ 1996b).
The frequency dependence of this emission,
rising sharply at long wavelengths,
is inconsistent with the expected microwave dust emission.
A 2-component fit to dust plus radio emission,
$$
I_\nu ~= ~A_{\rm dust} \left( \frac{\nu}{\nu_0} \right)^{\beta_{\rm dust}} 
        B_\nu(T_{\rm dust})
 ~+ ~A_{\rm radio} \left( \frac{\nu}{\nu_0} \right)^{\beta_{\rm radio}}
$$
with $1.5 < \beta_{\rm dust} < 2$ 
yields spectral index 
$\beta_{\rm radio} = -0.3^{+1.3}_{-3.8}$
for the unknown component,
strongly suggestive of free-free emission ($\beta_{\rm ff} = -0.15$).
At high latitudes ($|b| > 20\deg$),
spatial fluctuations in the diffuse synchrotron emission 
are uncorrelated with dust emission,
leaving free-free emission (thermal bremsstrahlung)
from the WIM as the only plausible alternative:
on large angular scales,
at least 30\% of microwave emission from the WIM is traced by
the diffuse infrared cirrus
(Kogut et al.\ 1996b).

The detection of correlation between dust and ionized gas at high latitudes is 
consistent with the correlation observed between dust and free-free emission
along the Galactic plane 
(Broadbent, Haslam, \& Osborne 1989).
Two questions of interest for cosmological observations are
the extent to which infrared dust 
reliably traces high-latitude radio emission,
and whether the correlation depends significantly on angular scale.
H$\alpha$ emission is a widely used tracer of the WIM.
Recently, 
McCullough 1997 
demonstrated a correlation between H$\alpha$ emission and
the IRAS infrared cirrus at 1\deg ~angular scale,
with an amplitude comparable to the value derived from DMR at much larger
angular scales.
In this paper, I present results from cross-correlations of
high-latitude H$\alpha$ maps
with the DIRBE 100 $\mu$m intensity 
on angular scales 0\ddeg7 to 10\deg.

\section{Analysis}
If H$\alpha$ emission is correlated with far-infrared dust,
the H$\alpha$ intensity in each pixel may be written as
\begin{equation}
H_i = a D_i + n_i,
\label{corr_eq}
\end{equation}
where 
$H$ is the intensity in the H$\alpha$ map,
$D$ is the far-infrared intensity in the dust map,
$n$ represents an uncorrelated component
(noise, systematic artifacts, or uncorrelated emission features),
and $i$ is a pixel index.
The presence of correlation between two data sets is often quantified
using the Pearson correlation coefficient,
$$
\rho = \frac{\sum_i (D_i - \langle D_i \rangle) (H_i - \langle H_i \rangle) }
	 { \left( \sum_i (D_i - \langle D_i \rangle)^2 
		  \sum_i (H_i - \langle H_i \rangle)^2 \right)^{1/2} } .
$$
While this has the advantage of being a simple, dimensionless quantity
that does not require detailed knowledge 
of noise or artifacts in the two data sets, 
it can not show whether the physically more interesting
correlation slope $a$ varies between different data sets
at different locations or angular scales.
An alternative approach uses the correlation slope $a$,
defined as
\begin{equation}
a = \frac{ \sum_{i,j} H_i ({\bf M}^{-1})_{ij} D_j }
         { \sum_{i,j} D_i ({\bf M}^{-1})_{ij} D_j }
\label{coeff_eq}
\end{equation}
with statistical uncertainty
$$
\delta a = \left[ \sum_{i,j} D_i ({\bf M}^{-1})_{ij} D_j \right]^{-1/2} ,
$$
where ${\bf M}$ is the covariance matrix for emission
between pixels $i$ and $j$ in the H$\alpha$ map.
In the limit ${\bf M}_{ij} = \sigma^2 \delta_{ij}$,
where $\sigma$ is the instrumental noise,
equation \ref{coeff_eq} reduces to a least-squares fit.
Note that for totally uncorrelated maps
$ \langle H D \rangle = 0$
averaged over many pixels (or many maps),
so that $\langle a \rangle = 0$ as expected when no correlation is present.
Equation \ref{coeff_eq} has the additional desirable property
of a well-defined distribution:
given two uncorrelated data sets $x$ and $y$
with unit normal distribution,
the distribution $a / \delta a$ will also 
have zero mean and unit standard deviation.
The presence of a true correlation slope $a_0$ in the parent population
shifts the mean fitted value $\langle a \rangle = a_0$
but does not alter the width of the distribution.

\single12
\begin{figure}[t]
\centerline{
\psfig{file=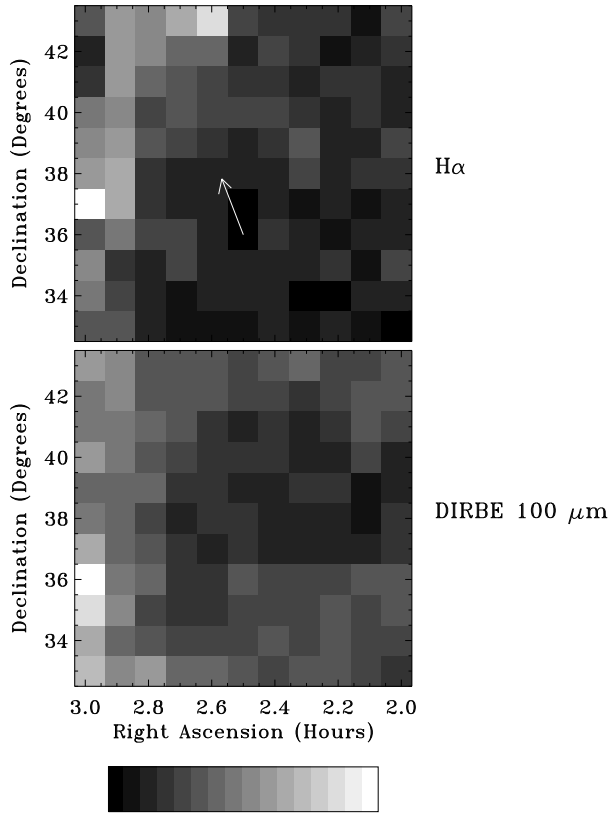,width=4.0in}}
\caption{Reynolds 1980 H$\alpha$ map and DIRBE 100 $\mu$m map 
in same sparse sampling.  
Full scale is 1--12 Rayleighs in H$\alpha$ 
and 3.7--10.3 MJy sr$^{-1}$ for DIRBE,
with white regions brighter for both maps.
The arrow shows the direction of lower Galactic latitude.}
\label{reynolds_dirbe_fig}
\end{figure}

\begin{figure}[t]
\centerline{
\psfig{file=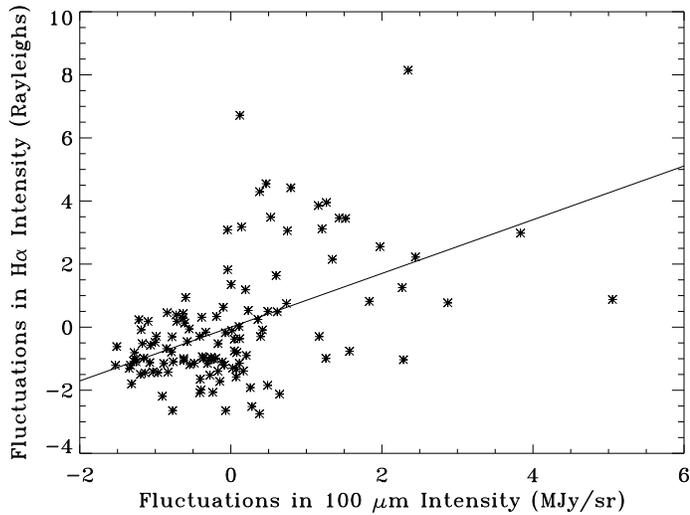,width=4.0in,angle=90}}
\caption{Reynolds 1980 H$\alpha$ map and DIRBE 100 $\mu$m map 
plotted point by point.
The fitted slope $a_0 = 0.85$ Rayleighs MJy$^{-1}$ sr.}
\label{tt_plot}
\end{figure}

\half12
Few measurements of diffuse H$\alpha$ emission exist 
at high Galactic latitude.  
Figure \ref{reynolds_dirbe_fig} shows 
the H$\alpha$ intensity within a $10\deg \times 12\deg$ patch
at $(l,b) = (144,-21)$
measured by a Fabry-Perot spectrometer 
with a 0\ddeg8 beam sampled on a 1\deg ~grid
(Reynolds 1980).
The 100 $\mu$m emission measured by DIRBE 
at 0\ddeg7 angular resolution
is shown in the same sparse sampling.
\footnote{Throughout this paper I use the 100 $\mu$m map
for its high signal to noise ratio and to facilitate
comparison with the IRAS 100 $\mu$m survey at finer angular scales.
I obtain similar results using the DIRBE 140 $\mu$m and 240 $\mu$m surveys.}
Figure \ref{tt_plot} plots the two maps point by point
and shows the fitted correlation slope $a$ for uniform pixel weight.
Quantification of the correlation slope 
between the two data sets
suffers from two problems.
Uncertainties in the subtraction of geocoronal emission
can contribute systematic artifacts comparable to
the instrumental noise in each line of sight 
(Reynolds 1980).
The small number of independent pixels (121) 
creates appreciable sample variance.
The far-IR dust emission has a steep power spectrum,
$P \propto \ell^{-3}$ where 
$\ell \propto \theta^{-1}$ is the spherical harmonic order
(Gautier et al.\ 1992, 
Kogut et al.\ 1996a).
Gradients in the H$\alpha$ map may thus correlate with large-scale features
in the far-IR dust emission simply by chance alignment.

\single12

\half12
These problems may be addressed using Monte Carlo techniques,
repeating the fit in equation \ref{coeff_eq}
after replacing the DIRBE patch $I_{\rm DIRBE}$
with a set of ``control'' patches 
selected from distant portions of the sky.
Since the same H$\alpha$ map is used for all trials,
the distribution of fitted slopes automatically accounts for
instrumental noise, systematic artifacts, and chance alignment,
and provides a simple, robust description of the probability 
for an uncorrelated patch of cirrus to produce a slope $a$
as large (or larger) than the value from the correct part of the sky.
Note that Eq. \ref{coeff_eq} is only asymptotically unbiased:
although the Monte Carlo technique correctly accounts 
for the width of the distribution in $a$,
instrumental noise in the ``template'' DIRBE map 
reduces the mean value by a factor
$S^2/(S^2 + n^2)$,
where $S$ and $n$ are the signal and noise in the DIRBE map.
For the data in this paper, $S/n \gg 1$
and the correction is negligible (less than 1\%).

Far-IR emission is markedly anisotropic,
so I restrict the control patches to those lying within the
latitude range of the H$\alpha$ map.
For computational ease,
I center each control patch 
on the pixel centers of the COBE quadrilateralized sky cube 
at resolution index 4 within the range
$15\deg < |b| < 30\deg$,
resulting in 64 independent cross-correlations.
To allow for possible correlated $\csc(b)$ emission 
(with no smaller-scale correlations),
I determine the distribution of fitted slope $a$
with the control patches in random orientation 
with respect to the H$\alpha$ map,
and again with the control patches fixed with respect to
the Galactic plane.

\single12
\begin{figure}[b]
\centerline{
\psfig{file=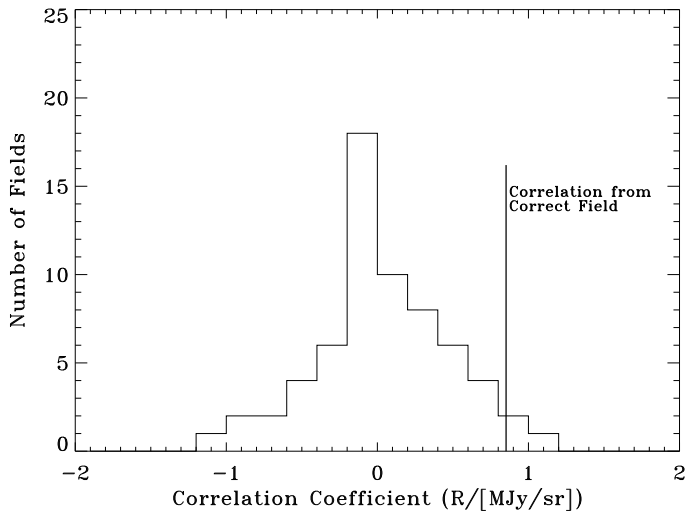,width=4.0in,angle=90}}
\caption{Histogram of fitted correlation slope between the 
Reynolds 1980 H$\alpha$ map and a set of 64 DIRBE ``control'' fields.
The correlation $a_0$ between the H$\alpha$ map and the DIRBE intensity
along the same lines of sight is also shown.}
\label{reynolds_histogram}
\end{figure}

\half12
Figure \ref{reynolds_histogram}
shows the distribution of the fitted correlation slopes $a$
between the Reynolds H$\alpha$ map 
and the set of randomly oriented control patches.
The correlation slope $a_0$ between the H$\alpha$ map
and the DIRBE 100 $\mu$m map in the correct position
is also shown.
Only 3\% of the uncorrelated control fields had $a > a_0$,
suggesting a marginal detection of correlated emission 
between H$\alpha$ and 100 $\mu$m dust,
$a_0 = 0.85 \pm 0.44$ Rayleighs MJy$^{-1}$ sr
within the Reynolds field.
Fixing the orientation of the control patches with respect to the Galactic 
plane does not appreciably alter the results:
the distribution of control patches still has zero mean and similar width,
demonstrating that the observed correlation
is not purely an artifact of general $\csc(b)$ trends in both data sets.
Repeating the analysis with the Pearson correlation coefficient $\rho$
instead of the correlation slope $a$ provides similar, slightly more 
significant, results:
$\rho = 0.48 \pm 0.22$,
significant at 98\% confidence.

\single12
\begin{figure}[b]
\centerline{
\psfig{file=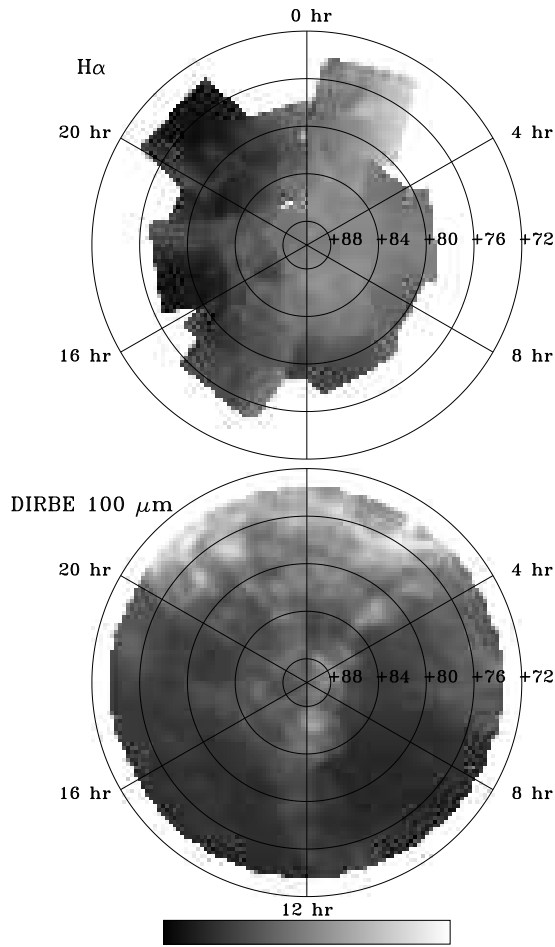,width=4.0in}}
\caption{H$\alpha$ mosaic over north celestial pole 
(Gaustad et al.\ 1996)
and fluctuations in DIRBE 100 $\mu$m intensity.
A mean has been removed from each data set.
White is brighter for both maps.}
\label{gaustad_dirbe_fig}
\end{figure}

\half12
Gaustad, McCullough, \& Van Buren 1996 
measured H$\alpha$ emission at 0\ddeg1 angular resolution
using narrow-band CCD images 
in a series of $7\deg \times 7\deg$ fields 
covering the north celestial polar cap 
(declination $\delta > 81\deg$).
Visible structure within the H$\alpha$ fields
is dominated by systematic artifacts (ghosts and stellar residuals),
with intrinsic anisotropy less than 1.3 Rayleighs 
(Gaustad, McCullough, \& Van Buren 1996).
To reduce systematic artifacts,
I reject all pixels in each field
lying more than 3 standard deviations from the mean in that field,
and then re-bin the remaining data 
into DIRBE pixels. 
Figure \ref{gaustad_dirbe_fig} shows a mosaic of
the re-binned H$\alpha$ map,
along with the DIRBE 100 $\mu$m data.

Artifacts associated with field edges
are evident in the H$\alpha$ mosaic,
and direct field to field comparison 
in regions where 2 or more fields overlap
confirms that much of the large-scale structure is systematic in origin.
Since power in the far-IR emission also increases 
on larger angular scales, 
the presence of large-scale systematics may be partially offset
by increased sensitivity to diffuse structure in the DIRBE maps.
To allow for this possibility, 
I cross-correlate the Gaustad et al.\ H$\alpha$ data
against the DIRBE data using two techniques.
The first method simply correlates the full H$\alpha$ mosaic
with the DIRBE 100 $\mu$m map,
using a set of control fields at $20\deg < |b| < 33\deg$
to estimate the uncertainty.
The full mosaic shows no statistically significant correlation:
28\% of the control fields had a correlation slope larger
than the actual fields in the correct orientation,
$a_0 = 0.21 \pm 0.43$ Rayleighs MJy$^{-1}$ sr.

The correlation from the full mosaic 
is dominated by systematic artifacts at the field edges.
A second method avoids these artifacts
by correlating each of the sixteen
$7\deg \times 7\deg$ fields independently,
then using the arithmetic mean 
of the 16 fitted correlation slopes $a_i$
to estimate the mean correlation over the full polar cap.
As before, a set of control fields provides an estimate 
for the uncertainty in the mean correlation.
17\% of the control fields had mean correlation larger
than the actual fields in the correct orientation,
yielding a correlation
$a_0 = 0.34 \pm 0.33$ Rayleighs MJy$^{-1}$ sr
on angular scales 0\ddeg7--7\deg 
~within the north celestial polar cap.
The results do not change appreciably if the control patches are held
in fixed orientation with respect to the Galactic plane.

\section{Discussion}
\begin{figure}[t]
\centerline{
\psfig{file=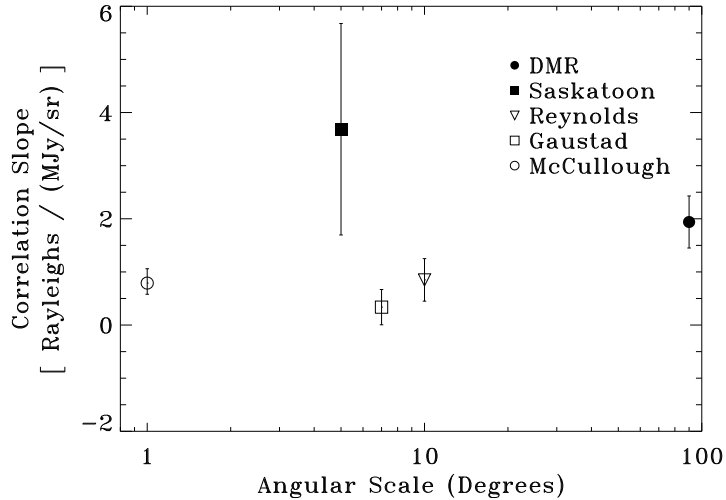,width=4.0in,angle=90}}
\caption{Correlation slopes $a$ 
between the warm ionized interstellar medium 
and far-infrared dust emission 
as a function of largest angle probed by each experiment.
Open symbols refer to H$\alpha$ data
and filled symbols refer to microwave measurements.
DMR: Kogut et al.\ 1996b;
Saskatoon: de Oliveira-Costa et al.\ 1997;
McCullough: McCullough 1997;
all others:this work.}
\label{correlation_vs_scale}
\end{figure}

Figure \ref{correlation_vs_scale}
plots the correlation slope between the WIM and 
100 $\mu$m dust emission
as a function of angular scale.
Since the dust power spectrum varies as $\theta^3$,
the data are shown at the largest angular scale probed by each experiment.
Data from microwave experiments
(DMR:Kogut et al.\ 1996b;
Saskatoon:de Oliveira-Costa et al.\ 1997)
are expressed in Rayleighs
assuming a spectral index $\beta_{\rm ff} = -0.15$
and a conversion 2 $\mu$K/Rayleigh at 53 GHz
appropriate for gas at 8000 K. 
There is no compelling evidence 
for variation in the correlation slope
as a function of angular scale.

\single12

\half12
The limited data in Figure \ref{correlation_vs_scale}
appear to fall in two groups, 
with correlations derived from microwave observations
lying higher than direct H$\alpha$ correlations,
even in the NCP field 
where both microwave and H$\alpha$ data are available.
Although the uncertainties on any individual measurement are appreciable,
the mean slope derived from microwave observations
lies 2.7 standard deviations 
above the mean of the three direct H$\alpha$ observations.
Both observations detect structure correlated with the spatial distribution
of the infrared cirrus,
but the microwave experiments appear to observe more emission per unit dust 
than are traced in H$\alpha$. 

Further evidence for ``extra'' microwave emission 
traced by the infrared cirrus comes from 
the MAX5 microwave anisotropy experiment
(Lim et al.\ 1996).
Data from the MAX5 experiment
show statistically significant correlations
between the IRAS 100 $\mu$m map
and millimeter-wavelength anisotropy measurements centered at
at 105, 180, 270, and 420 GHz.
The frequency dependence is consistent with a superposition of dust 
plus a second component identified as either
free-free emission 
or anisotropy in the cosmic microwave background;
although free-free emission is preferred,
the data do not distinguish decisively
between these alternatives.
The amplitude of the second component is large:
if it results from free-free emission,
the implied correlation slope 
between the WIM and dust is 
$a = 37 \pm 19$  Rayleighs MJy$^{-1}$ sr
on angular scales $\theta \approx 2\deg$
within the MAX5 $\mu$ Pegasi field
(Lim et al.\ 1996).
Such a result would imply even greater variation
in WIM/dust correlation than already evident 
in Figure \ref{correlation_vs_scale}
(since the MAX5 result can not be attributed definitively
to Galactic emission, it is not plotted).
Variation in the electron temperature within the WIM
would change the ratio of microwave free-free emission to H$\alpha$ emission.
However, the temperature of the high-latitude cirrus
(presumably heated by the same interstellar radiation field as the WIM)
is constant within 10\% when averaged over broad regions 
(Reach et al.\ 1995),
making problematic any preferential heating of the WIM.
Other possible emission mechanisms 
that would produce long-wavelength emission 
correlated with emission from the IR cirrus
include flat-spectrum synchrotron from supernovae shock fronts
and radio emission from rotating charged dust in the WIM
(Ferrara \& Dettmar 1994).

\section{Summary}
Cross-correlation of the DIRBE 100 $\mu$m survey
with previously published H$\alpha$ maps yields
positive correlation with 
amplitude comparable to previously published 
H$\alpha$ and microwave results.
The combined data show no dependence on angular scale, 
indicating that existing infrared maps can be used 
to trace microwave emission
over scales 0\ddeg7 ~to 90\deg.
Microwave experiments show a marginally significant trend 
toward observing more microwave emission 
per unit dust emission
than are traced by the same structures observed in H$\alpha$.
Additional observations, both H$\alpha$ and microwave,
are required to determine if this results from a statistical fluke,
variations within the WIM,
or the existence of competing emission mechanisms 
traced by the spatial distribution of high-latitude dust.

\vspace{0.5 in}
Gary Hinshaw provided helpful conversations.
This work was supported in part by NASA RTOP 399-20-61-01.

\clearpage

\begin{center}
\large
References
\end{center}

\normalsize
\half12

\refitem
Banday, A.\ \& Wolfendale, A.W.\ 1991, MNRAS, 248, 705

\refitem
Bennett, C.L., et al.\ 1992, ApJ, 396, L7

\refitem
Bennett, C.L., et al.\ 1994, ApJ, 434, 587

\refitem
Broadbent, A., Haslam, C.G.T., \& Osborne, J.L.\ 1989, MNRAS, 237, 381

\refitem
de Oliveira-Costa, A., et al.\ 1997, ApJ, 482, L17

\refitem
D\'{e}sert, F.-X., Boulanger, F., \& Puget, J.-L.\ 1990, A\&A, 327, 215 

\refitem
Ferrara, A., \& Dettmar, R.-J.\ 1994, ApJ, 427, 155

\refitem
Gaustad, J.E., McCullough, P.P, \& Van Buren, D.\ 1996, PASP, 108, 351

\refitem
Gautier, T.N., Boulanger, F., P\'{e}rault, M., \& Puget, J.L.\ 1992,
AJ, 103, 1313

\refitem
Kogut, A., et al.\ 1996a, ApJ, 460, 1

\refitem
Kogut, A., et al.\ 1996b, ApJ, 464, L5

\refitem
Lim, M.A., et al.\ 1996, ApJ, 469, L69

\refitem
McCullough, P.R.\ 1997, AJ, 113, 2186

\refitem
Reach, W.T., et al.\ 1995, ApJ, 451, 188

\refitem
Reynolds, R.J.\ 1980, ApJ, 236, 153

\refitem
Reynolds, R.J.\ 1992, ApJ, 392, L35

\end{document}